\documentclass[10pt,a4paper]{article}
\usepackage{amsmath,amssymb,amsfonts}
\usepackage{graphicx}
\usepackage{hyperref}
\usepackage[margin=0.75in]{geometry}
\usepackage{float}
\usepackage{bm}
\usepackage{booktabs}
\usepackage{tabularx}
\usepackage{subcaption}

\usepackage{setspace}
\usepackage{titlesec}
\titlespacing*{\section}{0pt}{8pt}{4pt}
\titlespacing*{\subsection}{0pt}{6pt}{3pt}

\AtBeginDocument{
  \setlength{\abovedisplayskip}{4pt}
  \setlength{\belowdisplayskip}{4pt}
  \setlength{\abovedisplayshortskip}{2pt}
  \setlength{\belowdisplayshortskip}{2pt}
}

\let\oldbibliography\thebibliography
\renewcommand{\thebibliography}[1]{%
  \oldbibliography{#1}%
  \setlength{\itemsep}{0pt}%
  \setlength{\parskip}{0pt}%
}

\newcommand{\dd}{\mathrm{d}}

\title{An Idealized Model of Buoyant Gas Spreading and Silt Release beneath Rough Cave Ceilings}

\author{
Sandy Hardian Susanto Herho$^{1,2,3,*}$,
Dasapta Erwin Irawan$^{1}$,\\ Rusmawan Suwarman$^{4}$, and
Deny Juanda Puradimaja$^{1}$
}

\date{}

\begin{document}
\maketitle

\begin{center}
\small
$^{1}$Applied Geology Research Group, Bandung Institute of Technology,
Bandung, West Java 40132, Indonesia\\
$^{2}$Department of Earth and Planetary Sciences, University of California, Riverside, CA 92521, USA\\
$^{3}$Indonesian Subaquatic Sport Association (POSSI), North Jakarta, DKI Jakarta 14240, Indonesia\\
$^{4}$Atmospheric Science Research Group, Bandung Institute of Technology,
Bandung, West Java 40132, Indonesia\\
$^{*}$Correspondence: sh001@ucr.edu
\end{center}

\begin{abstract}
\noindent
Gas exhaled by open-circuit divers collects under the ceilings of
submerged caves, migrates along the rock, and dislodges loose silt that then
falls through the passage, a process that cave divers call percolation.
This study develops an idealized model of the process. Buoyant gas under a
self-affine rough ceiling is treated as a quasi-static sequence of pools that
fill to their spill points, release gas that climbs to the next high point,
and merge where their levels meet. Moving contact lines strip silt from the
rock, and the released silt settles as a spectrum of grain sizes that
reduces the distance at which a diver can see. The computation agrees
exactly with an independent depression-filling algorithm, conserves gas
volume to round-off, and reproduces the known exponents of invasion
percolation in its capillary limit. A balance between the relief of the
ceiling and its dip defines a pool scale that organizes where gas collects,
and the area covered by gas from a single source grows more slowly than the
volume supplied, as predicted by the self-affine geometry. The cutoff at the
pool scale is soft, so large pools persist on steeper ceilings than the
scale alone would suggest. In a reference passage, the lead diver of a team
releases far more silt than the diver behind, and a team
releases far less than the sum of its members acting alone, because the
lead diver fills the pools and strips the paths that later divers reuse.
This ordering holds across every detachment efficiency and ceiling geometry
tested. Visibility toward the exit at eye level continues to decline for an
hour or more after a team has passed, as fine silt descends from a nearly
opaque layer beneath the ceiling. Absolute silt masses and sighting ranges
depend on quantities that have not been measured in caves, and the results
are best read as scaling relations and comparisons between scenarios.
\end{abstract}


\section{Introduction}
\label{sec:intro}

Flooded caves are among the most demanding environments in recreational
and scientific diving. Karst aquifers host extensive conduit networks in
which passages may run for kilometers below the water table, with walls,
floors, and ceilings shaped by dissolution and coated with fine sediment
deposited by slow flow \cite{FordWilliams2007}. A diver in such a passage has
no direct ascent to the surface, carries a finite supply of gas, and must
retrace the route in. The fatality record of cave diving has been reviewed
systematically \cite{Buzzacott2009}, and much of the training that cave
divers receive is devoted to preserving visibility and to following a
continuous guideline to the exit. Visibility in a cave can change within
seconds. A careless fin stroke near the floor raises a cloud of silt, and
in still water such a cloud may persist for long periods, because the
finest particles settle very slowly and little current exists to carry them
away \cite{DaviesColley2001}.

A second source of lost visibility originates on the ceiling rather than the
floor. Gas exhaled through an open-circuit regulator rises as a stream of
bubbles, spreads beneath the rock, collects in high points, and moves along
the ceiling as further gas arrives. Where the gas-water interface sweeps
across the rock, it loosens silt, clay, and small flakes that then fall into
the passage. Cave divers refer to this as percolation, and the term is used
in practice to explain why a passage that appeared clear behind a team can
be turbid on the return. The phenomenon is widely recognized among divers
but, as far as we are aware, it has not been treated quantitatively. The
questions it raises are practical and physical at the same time: where the
exhaled gas goes on a rough, inclined ceiling, how much silt it releases,
how that depends on the number of divers and their breathing, and how long
the released silt affects visibility along the route out.

The motion of gas under a rough ceiling belongs to a class of problems that
has been studied extensively in porous media. When a nonwetting fluid
displaces a wetting one slowly, capillary forces control the advance, and
the invading fluid enters the accessible pore of lowest entry pressure at
every step. Wilkinson and Willemsen \cite{WilkinsonWillemsen1983} showed that
this rule, invasion percolation, produces a critical percolation cluster.
The phase diagram of immiscible displacement, organized by the capillary
number and the viscosity ratio, places slow gas invasion in the regime of
capillary fingering \cite{Lenormand1988}, and the advance of the front
proceeds through abrupt pore-scale jumps rather than smoothly
\cite{Maloy1992}. Gravity modifies this picture. A density difference adds a
hydrostatic gradient to the entry thresholds \cite{Wilkinson1984}, which
stabilizes or destabilizes the front depending on its direction. In the
destabilizing case, a buoyant nonwetting fluid rises in a single finger
whose width decreases with the gradient as a power law fixed by the
exponents of percolation \cite{Birovljev1991, Meakin1992}, and the rising
fluid can break into clusters that migrate and fragment \cite{Wagner1997}.

A cave ceiling differs from a porous medium in that its relief is resolved
at the scale of interest, centimeters to meters, rather than hidden in pore
geometry. Buoyant gas under such a ceiling behaves like water on a
landscape turned upside down. Water on a land surface fills depressions,
spills over their lowest outlet, and joins depressions downslope, a
fill-and-spill behavior first described in hydrological field studies
\cite{SpenceWoo2003} and since formalized as an exact hierarchy of filling,
spilling, and merging for routing flow over digital elevation models
\cite{Barnes2021}. The same hierarchy, applied to the inverted ceiling,
describes how gas pools fill and overflow. The statistics of the relief
then matter. Natural rock surfaces, including fracture and fault surfaces,
are commonly self-affine over many decades of length, with a Hurst exponent
that sets how relief grows with scale \cite{Candela2012}. Combined with the
mean dip of a passage, self-affine relief introduces a competition between
roughness, which creates closed high points, and inclination, which removes
them.

The release of silt is controlled by contact lines. The edge of a gas pool
does not advance continuously over rock. Surface heterogeneity and
contact-angle hysteresis pin the contact line until the driving pressure
exceeds a threshold \cite{deGennes1985, JoannyDeGennes1984}, which adds a
capillary contribution to the purely gravitational picture. When the line
does move, it exerts a capillary force on attached particles that exceeds
their adhesion for particles larger than about a micrometer
\cite{Sharma2008}, and the efficiency of detachment depends on the shape of
the particles \cite{Aramrak2013}. Each passage of a gas-water interface can
therefore strip loose particles from rock.

Once released, the particles settle. Their settling velocity grows with the
square of their size in the Stokes regime and more slowly for larger grains
\cite{FergusonChurch2004}, and in saline or organic-rich water fine
particles may flocculate and settle faster than primary grains
\cite{Winterwerp2002}. Their effect on a diver depends on the attenuation of
light that they produce. Beam attenuation per unit mass is largest for the
finest particles \cite{Boss2009}, and the distance at which a dark target
remains visible is inversely proportional to the beam attenuation
coefficient \cite{Zaneveld2003}. Visibility is therefore governed by the
fine fraction, which settles most slowly. Lagrangian random-walk models are
a standard way to compute the vertical distribution of such particles in a
mixed water column \cite{Visser1997}.

This study links these elements in an idealized model of percolation. Its
aims are to identify which geometric and operational factors control how
much silt exhaled gas releases, where that silt originates relative to the
divers, and how it affects visibility along the route out of a cave. The
model does not attempt to predict the behavior of a particular passage. No
observational data are used. Every parameter is a round value within a
range we consider plausible, and the sensitivity of the results to the
quantities that have not been measured is reported alongside them. The
contributions are fourfold. We formulate the motion of exhaled gas under a
rough ceiling as an exact fill-spill-merge computation with capillary
pinning, and we verify it against an independent algorithm and known
results of percolation theory. We derive a pool scale and an area-volume
law from the self-affine geometry and test them against simulations. We
quantify how silt release depends on the position of a diver in a team, the
breathing rate, and holding station. Finally, we follow the released silt
through the water column to the sighting range at eye level on the route
out. The verification follows the practice adopted for related idealized
solvers \cite{Herho2025kh2d, Irawan2026amerta, Irawan2026kdv,
Herho2026wave}.

\section{Methods}
\label{sec:methods}

\begin{figure}[H]
    \centering
    \includegraphics[width=\linewidth]{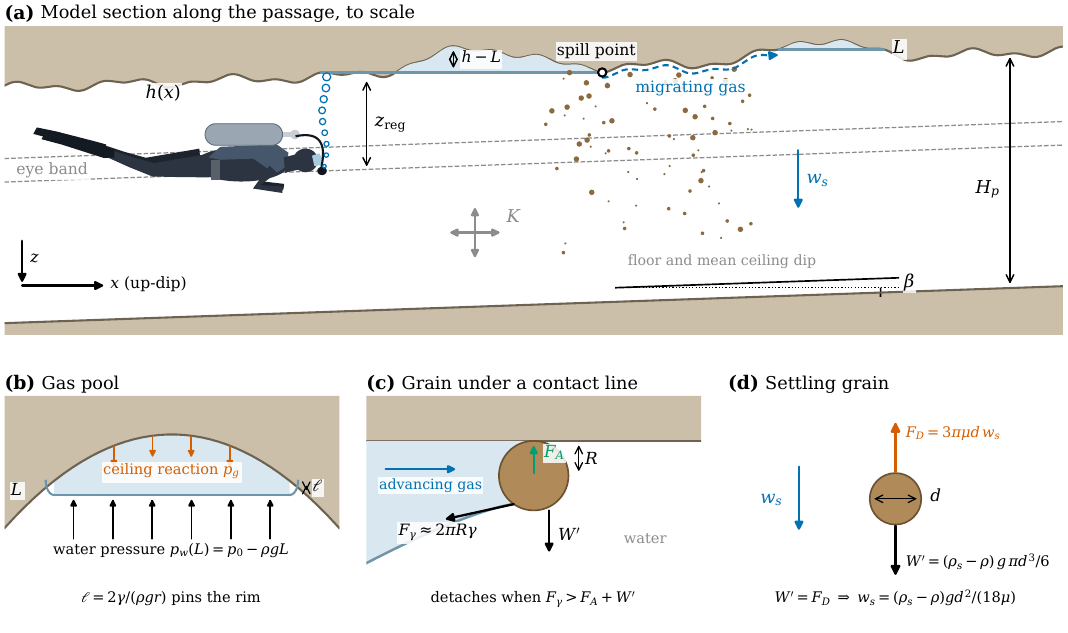}
    \caption{Model section and free-body diagrams. (a) Section along the
    passage, to scale. An idealized rough ceiling $h(x)$ with dip $\beta$
    holds gas pools with flat interfaces at levels $L$. The down-dip pool
    is full to its spill point, and gas crossing the saddle climbs along
    the ceiling to the next pool. An open-water diver exhales from the
    second stage at $z_{\rm reg}$ below the mean ceiling; silt stripped by
    moving contact lines settles at $w_{s}$ and spreads with eddy
    diffusivity $K$. (b) Gas pool. The water pressure $p_{w}(L)$ on the
    flat interface balances the ceiling reaction, which carries the gas
    pressure $p_{g}$; at the rim the contact line is pinned until the gas
    thickness exceeds $\ell$. (c) Loose grain of radius $R$ under an
    advancing contact line, with capillary force $F_{\gamma}$, adhesion
    $F_{A}$, and net weight $W'$. (d) Settling grain of diameter $d$, for
    which the net weight $W'$ balances the Stokes drag $F_{D}$ at the
    settling velocity $w_{s}$.}
    \label{fig:schematic}
\end{figure}

The configuration is shown in Fig.~\ref{fig:schematic}. A passage of length
12.8~m and width $W = 3.2$~m is bounded above by a rough ceiling and below
by a floor at a distance $H_{p} = 2$~m beneath the mean ceiling. The
coordinate $x$ runs along the passage and is positive up-dip, $y$ runs
across it, and $z$ is depth, measured downward, so that the elevation of a
point is $-z$. Divers swim along the centerline with their second stage at
$z_{\rm reg} = 0.8$~m below the mean ceiling, and visibility is evaluated in
an eye band between 0.6 and 0.8~m below it. All quantities are in SI units
unless stated otherwise: lengths in m, times in s, pressures in Pa, forces
in N, gas volumes in m$^{3}$ (quoted in liters where convenient), silt mass
in g, and silt concentration in g\,m$^{-3}$, which equals mg\,L$^{-1}$.
Throughout, the symbol $\equiv$ introduces a definition and $=$ denotes an
equality that is derived or imposed as a model relation.

The ceiling elevation $h(x,y)$ (m), positive upward, is the sum of a random
relief $h_{r}$, a uniform rise along the passage at the dip angle $\beta$,
and a cross-passage arch of height $A$ (m),
\begin{equation}
    h(x,y) = h_{r}(x,y) + x\tan\beta - A\left(\frac{2y}{W}-1\right)^{2}.
    \label{eq:ceiling}
\end{equation}
The arch term vanishes on the centerline $y = W/2$ and equals $-A$ at the
walls, so the ceiling is highest along the axis of the passage, as in a
phreatic tube. The relief is written as a Fourier integral,
$h_{r}(\mathbf{x}) = \int \hat{h}(\mathbf{k})\, e^{\mathrm{i}\mathbf{k}
\cdot\mathbf{x}}\,\dd^{2}k$, with independent Gaussian coefficients of
spectral density $\langle|\hat{h}(\mathbf{k})|^{2}\rangle = P(k)$ (m$^{4}$),
where $k = |\mathbf{k}|$ (m$^{-1}$). The mean square height difference over
a lag $\mathbf{r}$ is then
\begin{equation}
    S^{2}(r) \equiv \left\langle \left[h_{r}(\mathbf{x}+\mathbf{r})
      - h_{r}(\mathbf{x})\right]^{2}\right\rangle
      = 2\int P(k)\left[1 - \cos(\mathbf{k}\cdot\mathbf{r})\right]
        \dd^{2}k .
    \label{eq:structure_def}
\end{equation}
For a power-law spectrum $P(k) = P_{0}\,k^{-(2+2H)}$ with $0 < H < 1$, the
substitution $\mathbf{q} = r\mathbf{k}$ in Eq.~\eqref{eq:structure_def}
gives $S^{2}(r) = r^{2H}\,2P_{0}\int q^{-(2+2H)}[1 - \cos(q\cos\varphi)]
\,\dd^{2}q$, in which the remaining integral is a pure number that converges
at small $q$ because $1-\cos$ vanishes as $q^{2}$ and at large $q$ because
$H > 0$. The relief is therefore self-affine,
\begin{equation}
    S(r) = \sigma_{1}\left(\frac{r}{r_{0}}\right)^{H},
    \qquad r_{0} \equiv 1~\mathrm{m},
    \label{eq:structure}
\end{equation}
where $\sigma_{1}$ (m) is the root-mean-square height difference at a lag of
1~m and $H$ is the Hurst exponent. Because $H < 1$, the slope of the relief
over a lag $r$, of order $S(r)/r \propto r^{H-1}$, grows as $r$ decreases:
the ceiling is steep at small scales and gentle at large ones. We take
$H = 0.8$, within the range reported for natural rock surfaces
\cite{Candela2012}, and $\sigma_{1} = 0.10$~m.

The water is at rest. The vertical momentum balance then reduces to
$\dd p/\dd z = \rho g$, with $\rho = 1000$~kg\,m$^{-3}$ and
$g = 9.81$~m\,s$^{-2}$, which integrates to the hydrostatic pressure
\begin{equation}
    p_{w}(z) = p_{0} + \rho g z ,
    \label{eq:hydrostatic}
\end{equation}
in Pa, where $p_{0}$ is the pressure at the reference level $z = 0$. The
same balance applies inside the gas, with the gas density $\rho_{g}$ in
place of $\rho$. At a ceiling depth of 20~m, the gas is at about 3~bar and
$\rho_{g} \approx 3.6$~kg\,m$^{-3}$, so across a pool 0.2~m deep the gas
pressure changes by $\rho_{g}g\Delta z \approx 7$~Pa while the water
pressure changes by $\rho g\Delta z \approx 2000$~Pa. We therefore treat
the gas pressure $p_{g}$ as uniform within a connected pool. On scales large
compared with the capillary length, the pressure is continuous across the
gas-water interface, so the interface lies where
$p_{w}(z) = p_{g}$. By Eq.~\eqref{eq:hydrostatic}, this condition holds on a
single horizontal plane,
\begin{equation}
    z_{L} = \frac{p_{g} - p_{0}}{\rho g},
    \qquad L \equiv -z_{L},
    \label{eq:level}
\end{equation}
where $L$ (m) is the elevation of the interface, the level of the pool.
Fig.~\ref{fig:schematic}b shows the forces on the gas of one pool. The
water pushes up on the flat interface with the pressure
$p_{w}(z_{L}) = p_{g}$, and the rock pushes down on the gas with the same
pressure, so the gas is in equilibrium for any shape of the ceiling. The
force that the gas exerts on the rock in excess of the hydrostatic water
pressure that would act there without gas is obtained by integrating the
pressure difference over the horizontal projection of the covered ceiling,
$\int [p_{g} - p_{w}(-h)]\,\dd A = \int \rho g\,(h - L)\,\dd A$. With
Eq.~\eqref{eq:volume_cont} below, this force is $\rho g V$, the buoyancy of
the pool, which the ceiling carries.

At the rim of a pool, the contact line does not advance onto rock as soon as
the interface reaches it. Micro-roughness, cracks, and the hysteresis of the
contact angle pin the line \cite{deGennes1985, JoannyDeGennes1984}, and gas can only enter a crevice of radius $r$
(m) when the pressure difference across the meniscus overcomes the
Young-Laplace pressure. For a water-wet surface with two principal
curvatures of order $1/r$, this pressure is $2\gamma/r$, with
$\gamma = 0.072$~N\,m$^{-1}$ the surface tension. At a point of the ceiling
at elevation $h$ under a pool of level $L$, the pressure difference between
gas and water is, by Eqs.~\eqref{eq:hydrostatic} and \eqref{eq:level},
$p_{g} - p_{w}(-h) = \rho g\,(h - L)$, the weight per unit area of a water
column of height $h - L$. The contact line can advance only if this
difference exceeds $2\gamma/r$, which gives
\begin{equation}
    h - L \geq \ell \equiv \frac{2\gamma}{\rho g\, r}
    \qquad\Longleftrightarrow\qquad
    L \leq h_{\rm eff} \equiv h - \ell .
    \label{eq:pinning}
\end{equation}
The pinning height $\ell$ has units of (N\,m$^{-1}$)/(N\,m$^{-3}$ $\cdot$ m)
$=$ m. Equation~\eqref{eq:pinning} states that the gas sees an effective
ceiling lowered by $\ell$: a cell is covered when the level of the pool
lies at or below $h_{\rm eff}$. The radius $r$ is drawn independently for
each cell from a log-normal distribution with median $r_{50} = 4$~mm and
logarithmic standard deviation 0.5, which gives a median pinning height
$\ell_{50} = 3.7$~mm, comparable to the capillary length
$\ell_{c} \equiv (\gamma/\rho g)^{1/2} = 2.7$~mm. The ratio of the relief
across one lattice cell of size $a$ to the pinning height,
$\mathrm{Bo} \equiv S(a)/\ell_{50} = 1.19$ for $a = 2$~cm, plays the role of
a Bond number: gravity acting through the relief and capillarity acting
through pinning are of comparable importance at the scale of one cell, and
gravity dominates at larger scales because $S$ grows with $r$.

The volume of gas held by a pool follows from the same picture. Below each
covered point of the ceiling, the gas occupies the column between the
effective ceiling and the level, so that
\begin{equation}
    V = \int_{\mathcal{A}} \left(h_{\rm eff} - L\right)\dd A ,
    \label{eq:volume_cont}
\end{equation}
where $\mathcal{A}$ is the covered area. The integrand is positive by
Eq.~\eqref{eq:pinning}, and $V$ has units of m$^{3}$. On a lattice of
spacing $a$, the midpoint rule over the $n$ covered cells $i$ of a pool
$\mathcal{P}$ gives
\begin{equation}
    V = a^{2} \sum_{i\in\mathcal{P}} \left(h_{{\rm eff},i} - L\right)
      = a^{2}\left(\Sigma_{\mathcal{P}} - nL\right),
    \qquad
    \Sigma_{\mathcal{P}} \equiv \sum_{i\in\mathcal{P}} h_{{\rm eff},i} .
    \label{eq:volume}
\end{equation}
Differentiating Eq.~\eqref{eq:volume} with respect to $L$ at fixed
$\mathcal{P}$ gives $\partial V/\partial L = -na^{2}$, so that a small
injection $\Delta V$ lowers the level by
\begin{equation}
    \Delta L = -\frac{\Delta V}{n a^{2}} .
    \label{eq:dlevel}
\end{equation}
Physically, added gas spreads over the whole covered area and pushes the
interface down uniformly. As $L$ descends, the first perimeter cell to
satisfy Eq.~\eqref{eq:pinning} is the one with the highest effective
ceiling, so the pool grows by admitting perimeter cells in decreasing order
of $h_{\rm eff}$. This is the invasion-percolation rule with threshold
$-h_{\rm eff}$ \cite{WilkinsonWillemsen1983}. A perimeter cell $j$ is a
spill point when one of its neighbors $m$ outside the pool lies higher,
$h_{{\rm eff},m} > h_{{\rm eff},j}$. At $L = h_{{\rm eff},j}$, gas reaching
$j$ is no longer confined: buoyancy drives it toward higher ceiling, which
lies outside the pool. The pool is then full. Any further gas crosses the
saddle at $j$ and rises along the path of steepest ascent of
$h_{\rm eff}$, $c_{k+1} = \arg\max_{m\in\mathcal{N}(c_{k})}
h_{{\rm eff},m}$, where $\mathcal{N}$ denotes the four lattice neighbors,
until it reaches a crest, where it starts a new pool, or a cell of an
existing pool, which it joins. When two pools meet at a common saddle with
equal levels, they merge into one. Gas that reaches either open end of the
passage leaves the modeled section. This is the fill-spill-merge
construction of landscape hydrology \cite{SpenceWoo2003, Barnes2021}, applied to the
inverted ceiling.

In this description, gas that crosses a saddle does not remain hydraulically
connected to the pool it left. A connected gas body would, by
Eq.~\eqref{eq:level}, have one level and would flood every cell above the
saddle elevation at once. Raising a gas layer by only 1~cm under 1~m$^{2}$
of ceiling requires 10~L, about seven breaths, so the volume delivered by a
breath cannot maintain such a connection. Gas that crosses a saddle instead
forms a thin stream that rises to the next crest, as in buoyancy-driven
migration with fragmentation \cite{Wagner1997}.

When every pool is full, the covered set no longer depends on the order in
which gas arrived, because each pool is filled to the elevation of its spill
point. That state is the filling of the depressions of
$f \equiv -h_{\rm eff}$ from the outlets at the two ends of the passage,
\begin{equation}
    F_{m} = \max\left(f_{m},\, F_{c}\right),
    \qquad
    h_{\rm eff} - L = F - f ,
    \label{eq:flood}
\end{equation}
where $F$ (m) is the filled surface and $c$ is the neighbor through which
cell $m$ is first reached in order of increasing $F$. Equation~\eqref{eq:flood}
provides an exact check on the event-driven computation of
Eqs.~\eqref{eq:volume} and \eqref{eq:dlevel}.

Each diver exhales, once per breathing period $T_{b}$ (s), a volume
$\mathrm{RMV}\,T_{b}/60$ at the ambient pressure of the regulator, where RMV
is the respiratory minute volume in L\,min$^{-1}$ and the factor 60 converts
minutes to seconds. Exhaled bubbles reach thermal equilibrium with the water
quickly, so their expansion between the regulator and the ceiling is
isothermal, $pV = \mathrm{const}$ (Boyle's law). With
Eq.~\eqref{eq:hydrostatic} evaluated at the regulator, at water depth
$D + z_{\rm reg}$, and at the ceiling, at water depth $D$, the volume per
breath at the ceiling is
\begin{equation}
    V_{b} = \frac{\mathrm{RMV}\,T_{b}}{60}\,
      \frac{p_{\rm atm} + \rho g\,(D + z_{\rm reg})}{p_{\rm atm} + \rho g D},
    \label{eq:breath}
\end{equation}
in liters, with $p_{\rm atm} = 101\,325$~Pa. For RMV = 20~L\,min$^{-1}$,
$T_{b} = 4$~s, $D = 20$~m, and $z_{\rm reg} = 0.8$~m, the pressure ratio is
1.027 and $V_{b} = 1.37$~L. The bubbles reach the ceiling in a Gaussian
footprint of root-mean-square radius 0.12~m centered above the regulator. A
closed-circuit rebreather exhales no gas in level swimming and is
represented by RMV = 0.

Silt is released by moving contact lines (Fig.~\ref{fig:schematic}c). A
particle of radius $R$ held at a gas-water interface experiences a capillary
force of maximum magnitude $F_{\gamma} \approx 2\pi R\gamma$ (N), the
surface tension acting along its line of contact. Its net weight in water is
$W' = (\rho_{s} - \rho)\,g\,\tfrac{4}{3}\pi R^{3}$, with
$\rho_{s} = 2650$~kg\,m$^{-3}$ for quartz-density grains. The ratio
\begin{equation}
    \frac{F_{\gamma}}{W'} = \frac{3\gamma}{2(\rho_{s}-\rho)\,g\,R^{2}}
    \label{eq:capweight}
\end{equation}
is dimensionless and equals about $7\times10^{4}$ for $R = 10~\mu$m, so
weight plays no role in detachment, and the outcome is decided by the
competition between $F_{\gamma}$ and the adhesion $F_{A}$ of the particle
to the rock. Experiments show that a moving air-water interface removes a
large fraction of attached particles larger than about a micrometer
\cite{Sharma2008}. Because the adhesion of silt on a cave ceiling is not
known, we represent the outcome by two efficiencies. When a contact line
crosses a cell for the first time, it removes the fraction $\phi_{c}$
(dimensionless) of the loose silt in that cell. When a volume $v$ (m$^{3}$)
of migrating gas crosses a cell, each bubble sweeps the cell anew, and we
take the remaining mass to decay as $m \to m\,e^{-v/v_{d}}$, with a
stripping volume $v_{d}$ (m$^{3}$). For a sequence of parcels $v_{k}$, the
remaining mass is multiplied by
\begin{equation}
    \prod_{k} e^{-v_{k}/v_{d}} = \exp\!\left(-\frac{1}{v_{d}}\sum_{k}
      v_{k}\right),
    \label{eq:strip}
\end{equation}
which depends only on the total volume, so the result does not depend on
how a breath is divided into parcels. The loose inventory is uniform,
$m_{0} = 50$~g\,m$^{-2}$, so each cell initially holds $m_{0}a^{2} =
0.02$~g, with $\phi_{c} = 0.5$ and $v_{d} = 2$~mL.

The geometry of the ceiling sets a scale for the pools. Over a horizontal
distance $\ell$ along the passage, the relief typically rises or falls by
$S(\ell)$ of Eq.~\eqref{eq:structure}, while the dip raises the ceiling by
$\ell\tan\beta$. A closed high point of width $\ell$ can hold gas only if
the relief can reverse the rise of the dip across it. Equating the two,
$\sigma_{1}(\ell/r_{0})^{H} = \ell\tan\beta$, and solving for $\ell$ gives
the pool scale
\begin{equation}
    r^{*} = r_{0}\left(\frac{\sigma_{1}}{r_{0}\tan\beta}\right)^{1/(1-H)},
    \label{eq:rstar}
\end{equation}
which has units of m because $\sigma_{1}/(r_{0}\tan\beta)$ is
dimensionless. Since $H < 1$, the relief dominates below $r^{*}$ and the dip
above it. Because the exponent $1/(1-H)$ is 5 for $H = 0.8$, $r^{*}$
changes by orders of magnitude over a few degrees of dip. The scale is not
a sharp cutoff. For Gaussian relief, the height difference across a width
$\ell$ is normally distributed with standard deviation $S(\ell)$, so the
probability that it exceeds the rise of the dip is
$\tfrac{1}{2}\,\mathrm{erfc}\!\left[\ell\tan\beta/(\sqrt{2}\,S(\ell))
\right]$. Using Eq.~\eqref{eq:rstar}, the argument satisfies
$[\ell\tan\beta/S(\ell)]^{2} = (\ell/r^{*})^{2(1-H)}$, and the leading
behavior of the complementary error function gives
\begin{equation}
    \Pr(\ell) \sim \exp\!\left[-\tfrac{1}{2}
      \left(\frac{\ell}{r^{*}}\right)^{2(1-H)}\right] .
    \label{eq:soft}
\end{equation}
This is a stretched exponential with exponent $2(1-H)$, which is 0.4 for
$H = 0.8$. It decays slowly, so pools several times larger than $r^{*}$
remain likely, and $r^{*}$ acts as a scaling variable rather than a limit.

Below $r^{*}$, the dip is negligible and the pools follow the self-affine
relief alone. A pool of width $\ell$ covers an area $A \sim \ell^{2}$ and is
filled to a depth set by the relief, $h - L \sim S(\ell) \propto \ell^{H}$.
By Eq.~\eqref{eq:volume_cont}, its volume is $V \sim A\,S(\ell) \propto
\ell^{2+H}$. Eliminating $\ell$ gives
\begin{equation}
    V \propto A^{(2+H)/2}
    \qquad\text{and}\qquad
    A \propto V^{2/(2+H)} .
    \label{eq:areavolume}
\end{equation}
The first relation applies across the pools of one ceiling. The second
applies to gas fed from a single point, whose pools fill and merge into a
cluster that remains self-similar as it grows. The exponent $2/(2+H)$ is
0.800 for $H = 0.5$ and 0.714 for $H = 0.8$. It is less than one, so each
additional liter of gas covers less new ceiling than the liter before it.

When the relief is negligible compared with the pinning heights, as on a
smooth, nearly flat ceiling, the level differences that drive
fragmentation vanish, the gas remains connected, and its contact line
advances into the perimeter cell with the lowest entry pressure. We
normalize the entry pressures $2\gamma/r_{i}$ to dimensionless thresholds
$t_{i}$ uniform on $(0,1)$. Water enclosed by gas on a ceiling remains
connected to the water below, so no region can be trapped, and the process
is invasion percolation without trapping \cite{WilkinsonWillemsen1983}. Its
cluster is a critical percolation cluster, whose mass grows with its radius
of gyration $R_{g}$ as $N \propto R_{g}^{D_{f}}$ with $D_{f} = 91/48$, and
the thresholds it accepts are bounded by the site percolation threshold
$p_{c} = 0.592746$ of the square lattice \cite{NewmanZiff2000}. A dip adds
the hydrostatic term to the thresholds \cite{Wilkinson1984}, $t_{i} = u_{i} - G\,y_{i}$, where
$y_{i}$ is the up-dip position in lattice units and the dimensionless
gradient $G = \rho g a\tan\beta/\Delta p_{c}$ is the change in hydrostatic
pressure per lattice step divided by the spread $\Delta p_{c}$ of entry
pressures. The finger width follows from a self-consistency argument. Near
$p_{c}$, percolation clusters are correlated over a length
$\xi \propto |p - p_{c}|^{-\nu}$, with $\nu = 4/3$ in two dimensions.
Across a finger of width $w$, the gradient shifts the thresholds by $Gw$,
which sets the local distance from $p_{c}$. Setting $w \sim \xi$ gives
$w \propto (Gw)^{-\nu}$, or
\begin{equation}
    w \propto G^{-\nu/(1+\nu)} = G^{-4/7},
    \label{eq:finger}
\end{equation}
in lattice units, the scaling established for gravity-destabilized invasion
\cite{Birovljev1991, Meakin1992}.

A released grain of diameter $d$ settles under its net weight
$W' = (\rho_{s}-\rho)\,g\,\pi d^{3}/6$ and the Stokes drag
$F_{D} = 3\pi\mu d\,w_{s}$, with $\mu = 1.0\times10^{-3}$~Pa\,s
(Fig.~\ref{fig:schematic}d). Its velocity relaxes to the terminal value
within a time of order $\rho_{s}d^{2}/(18\mu)$, less than 3~ms for the
largest grains, after which the two forces balance,
\begin{equation}
    (\rho_{s}-\rho)\,g\,\frac{\pi d^{3}}{6} = 3\pi\mu d\, w_{s}
    \qquad\Longrightarrow\qquad
    w_{s} = \frac{(\rho_{s}-\rho)\,g\,d^{2}}{18\mu},
    \label{eq:stokes}
\end{equation}
in m\,s$^{-1}$. Stokes drag assumes a particle Reynolds number
$\mathrm{Re}_{p} = w_{s}d\rho/\mu$ well below one. This holds for all
classes up to 64~$\mu$m, for which $\mathrm{Re}_{p} \le 0.24$, but not for
the coarsest class of 128~$\mu$m, with $\mathrm{Re}_{p} \approx 1.9$, whose
settling velocity is therefore overestimated \cite{FergusonChurch2004}. That class carries 0.4
percent of the mass and contributes little to attenuation.

The mass of each size class is conserved. In a width-averaged section of
the passage, its concentration $C(x,z,t)$ (g\,m$^{-3}$) changes only through
the divergence of its flux, $\partial_{t}C + \nabla\cdot\mathbf{J} = 0$. The
flux combines advection by the along-passage current $U$ (m\,s$^{-1}$),
settling at $w_{s}$, and turbulent mixing represented by Fick's law with an
eddy diffusivity $K$ (m$^{2}$\,s$^{-1}$),
$\mathbf{J} = (UC - K\partial_{x}C,\; w_{s}C - K\partial_{z}C)$. With
constant coefficients, this gives
\begin{equation}
    \frac{\partial C}{\partial t} + U\frac{\partial C}{\partial x}
      + w_{s}\frac{\partial C}{\partial z}
      = K\left(\frac{\partial^{2} C}{\partial x^{2}}
      + \frac{\partial^{2} C}{\partial z^{2}}\right).
    \label{eq:transport}
\end{equation}
No silt crosses the ceiling, so the vertical flux vanishes there,
$J_{z} = w_{s}C - K\partial_{z}C = 0$ at $z = 0$. Silt that reaches the
floor deposits and is not resuspended, $C = 0$ at $z = H_{p}$. The reference
case has $U = 0$ and $K = 10^{-5}$~m$^{2}$\,s$^{-1}$, which represents weak
residual mixing in still cave water.

The time that a grain spends in the water column sets how long it can
affect visibility. Let $T(z)$ (s) be the mean time for a grain at depth $z$
to reach the floor. Over a short interval $\delta t$ the grain moves by
$w_{s}\delta t$ plus a random displacement of variance $2K\delta t$, so
$T(z) = \delta t + \langle T(z + \delta z)\rangle$. Expanding to first order
in $\delta t$ gives the backward equation
\begin{equation}
    K\,\frac{\dd^{2}T}{\dd z^{2}} + w_{s}\frac{\dd T}{\dd z} = -1,
    \qquad
    \left.\frac{\dd T}{\dd z}\right|_{z=0} = 0,
    \qquad
    T(H_{p}) = 0,
    \label{eq:backward}
\end{equation}
where the first boundary condition expresses reflection at the ceiling and
the second absorption at the floor. With $u \equiv \dd T/\dd z$,
Eq.~\eqref{eq:backward} becomes $u' + (w_{s}/K)\,u = -1/K$, whose solution
with $u(0) = 0$ is $u = -(1 - e^{-w_{s}z/K})/w_{s}$. Integrating from $0$ to
$H_{p}$ and using $T(H_{p}) = 0$ gives, for a grain released at the
ceiling,
\begin{equation}
    T(0) = \frac{H_{p}}{w_{s}} - \frac{K}{w_{s}^{2}}
      \left(1 - e^{-\mathrm{Pe}}\right),
    \qquad
    \mathrm{Pe} \equiv \frac{w_{s}H_{p}}{K} .
    \label{eq:residence}
\end{equation}
The P\'eclet number Pe compares settling with mixing. For
$\mathrm{Pe} \gg 1$, $T(0) \to H_{p}/w_{s}$, the time to fall through the
passage. For $\mathrm{Pe} \ll 1$, expanding the exponential to second order
gives $T(0) \to H_{p}^{2}/(2K)$, the time to diffuse across it. The released
mass is divided among 22 size classes from 1 to 128~$\mu$m, spaced by a
factor $2^{1/3}$, according to a log-normal distribution with median
8~$\mu$m and logarithmic standard deviation 1. Their settling velocities
range from $9\times10^{-4}$ to 14.7~mm\,s$^{-1}$, and their residence
times from about 2~min to about two days.

Visibility is set by the attenuation of a light beam. A grain of diameter
$d$ removes light over a cross-section $Q_{c}\pi d^{2}/4$, where the
extinction efficiency $Q_{c}$ approaches 2 for grains much larger than the
wavelength. A mass concentration $C_{k}$ of class $k$ corresponds to
$C_{k}/(\rho_{s}\pi d_{k}^{3}/6)$ grains per m$^{3}$, with $\rho_{s}$
expressed in g\,m$^{-3}$. The beam attenuation coefficient is the product of
number concentration and cross-section,
\begin{equation}
    c = c_{w} + \sum_{k} \frac{C_{k}}{\rho_{s}\pi d_{k}^{3}/6}\,
      \frac{Q_{c}\pi d_{k}^{2}}{4}
      = c_{w} + \sum_{k} b_{k}C_{k},
    \qquad
    b_{k} = \frac{3}{\rho_{s}d_{k}},
    \label{eq:attenuation}
\end{equation}
in m$^{-1}$, where $c_{w}$ is the attenuation of clear water and $b_{k}$
(m$^{2}$\,g$^{-1}$) is the mass-specific attenuation. Because $b_{k}$ varies
as $1/d_{k}$, a given mass of fine silt attenuates light far more than the
same mass of coarse silt, consistent with the dependence of mass-specific
attenuation on particle size observed in natural waters \cite{Boss2009}. In uniform water, a black target is visible to a
distance $4.8/c$ \cite{Zaneveld2003}, which corresponds to an optical depth
$c\,y$ of 4.8. Along a path of nonuniform attenuation, the optical depth is
the integral of $c$, so we take the sighting range $y$ (m) toward the exit,
from a point $x_{0}$ looking down-dip, as the distance at which
\begin{equation}
    \int_{0}^{y} c(x_{0} - s)\,\dd s = 4.8 ,
    \label{eq:sighting}
\end{equation}
and we set $c_{w} = 0.16$~m$^{-1}$, so that the clear-water sighting range is
30~m. The reference parameters are listed in Table~\ref{tab:parameters}.

\begin{table}[H]
\centering
\caption{Reference parameters. All values are illustrative.}
\label{tab:parameters}
\small
\begin{tabular}{llll}
\toprule
Quantity & Symbol & Value & Unit \\
\midrule
Lattice spacing & $a$ & 0.02 & m \\
Passage length, width, height & -- & 12.8, 3.2, 2.0 & m \\
Hurst exponent, relief at 1~m & $H$, $\sigma_{1}$ & 0.8, 0.10 & --, m \\
Dip, arch & $\beta$, $A$ & 2, 0.15 & deg, m \\
Median pinning radius, log-SD & $r_{50}$ & 4, 0.5 & mm, -- \\
Respiratory minute volume & RMV & 20 & L\,min$^{-1}$ \\
Breathing period, exhalation & $T_{b}$ & 4.0, 1.6 & s \\
Swimming speed, spacing & -- & 0.2, 20 & m\,s$^{-1}$, s \\
Ceiling depth, regulator depth & $D$, $z_{\rm reg}$ & 20, 0.8 & m \\
Loose silt inventory & $m_{0}$ & 50 & g\,m$^{-2}$ \\
Detachment constants & $\phi_{c}$, $v_{d}$ & 0.5, 2 & --, mL \\
Median grain size, log-SD & $d_{50}$ & 8, 1.0 & $\mu$m, -- \\
Eddy diffusivity & $K$ & $10^{-5}$ & m$^{2}$\,s$^{-1}$ \\
Clear-water sighting range & $4.8/c_{w}$ & 30 & m \\
\bottomrule
\end{tabular}
\end{table}

The ceiling is discretized on a lattice of spacing $a = 2$~cm, with
$640 \times 160$ cells for the reference passage. The relief is generated by
spectral synthesis of Eq.~\eqref{eq:structure_def} on a lattice four times
larger in each direction and cropped, which removes the periodicity of the
discrete Fourier transform that would otherwise flatten $S(r)$ at lags
approaching the domain size. It is rescaled so that the mean of the two
lattice directions satisfies Eq.~\eqref{eq:structure} at $r = r_{0}$. Each
cell has four neighbors. The boundaries across the passage are closed, and
each cell at an open end has an additional exit node, placed one cell
beyond it with the elevation given by the extension of the dip.

The fill-spill-merge computation, \texttt{FillSpill}, is event driven. Each
pool stores $n$, $\Sigma_{\mathcal{P}}$, and $L$, so that
Eq.~\eqref{eq:volume} is evaluated in constant time, and a priority queue of
its perimeter ordered by $h_{\rm eff}$. An injection $\Delta V$ is applied
through Eq.~\eqref{eq:dlevel}. If the level would pass below the highest
perimeter cell, the volume needed to reach that cell, $(L -
h_{{\rm eff},j})\,na^{2}$, is consumed and the cell is either admitted or,
if it is a spill point, the remainder is routed along the precomputed
steepest-ascent path to the next pool. Merging appends the smaller pool to
the larger. Because every step conserves the sum of the pool volumes and the
volume leaving the section, the total volume is conserved to round-off.
Each breath of Eq.~\eqref{eq:breath} is divided into 40 parcels released at
random times during an exhalation of 1.6~s and at random points of the
footprint. The saturated state of Eq.~\eqref{eq:flood} is computed
independently by priority-flood filling, \texttt{priority\_flood}, which
shares no code with the event-driven computation. In the connected limit,
invasion percolation is computed with a binary heap on lattices of up to
$1201^{2}$ sites.

Equation~\eqref{eq:transport} is the Fokker-Planck equation of the
stochastic differential equations $\dd X = U\,\dd t + \sqrt{2K}\,\dd B_{x}$
and $\dd Z = w_{s}\,\dd t + \sqrt{2K}\,\dd B_{z}$, where $B_{x}$ and $B_{z}$
are independent Brownian motions. We therefore solve it with Lagrangian
particles, as is standard for particles in a mixed water column
\cite{Visser1997}. With constant coefficients, the increments over a step
$\Delta t$ are exactly Gaussian, and
\begin{equation}
    x^{n+1} = x^{n} + U\Delta t + \sqrt{2K\Delta t}\,\xi_{x},
    \qquad
    z^{n+1} = \left|z^{n} + w_{s}\Delta t + \sqrt{2K\Delta t}\,\xi_{z}
      \right| ,
    \label{eq:lagrange}
\end{equation}
where $\xi_{x}$ and $\xi_{z}$ are independent standard normal variables. The
absolute value reflects particles at the ceiling, which enforces the
no-flux condition, and particles that reach the floor are removed, which
enforces deposition. Detection of the floor at the ends of steps
introduces a bias of order $(\Delta t)^{1/2}$ in residence times, which we
measure against Eq.~\eqref{eq:residence}. The released mass of each size
class is represented by about $1.2\times10^{4}$ particles of equal mass,
with stochastic rounding so that the expected mass is exact. The
concentration is the particle mass in each bin divided by the bin volume,
$W\,\Delta x\,\Delta z$, with $\Delta x = 0.1$~m and $\Delta z = 0.05$~m,
and the step is $\Delta t = 2$~s. Equation~\eqref{eq:sighting} is evaluated
by cumulative summation of Eq.~\eqref{eq:attenuation} over the bins between
each point and the down-dip end of the section, with clear water beyond
it.

We verify the computation with five tests. The event-driven computation is
compared with Eq.~\eqref{eq:flood} on twelve ceilings with dips from 0 to
$8^{\circ}$, and its volume balance is recorded over 20\,000 random
injections. For a paraboloidal dome, $h = -\varrho^{2}/(2R)$ with $\varrho$
the horizontal distance from its apex and $R$ (m) its radius of curvature,
Eq.~\eqref{eq:volume_cont} gives a covered disk of area
$A = 2\pi R|L|$ and a volume $V = \pi R L^{2}$, so that
$A = 2\sqrt{\pi R V}$ exactly. This is compared with the computation at four
lattice spacings. The connected limit is compared with $D_{f}$ and $p_{c}$,
and the gradient limit with Eq.~\eqref{eq:finger}. The Lagrangian scheme is
compared with Eq.~\eqref{eq:residence} over three decades of Pe and three
step sizes. Ensembles use twelve ceiling realizations unless stated
otherwise, and confidence intervals are obtained by bootstrap resampling
over realizations. The code is written in Python with NumPy
\cite{Harris2020}, SciPy \cite{Virtanen2020}, and Matplotlib
\cite{Hunter2007}.

\section{Results}
\label{sec:results}

\begin{figure}[H]
    \centering
    \includegraphics[width=0.86\linewidth]{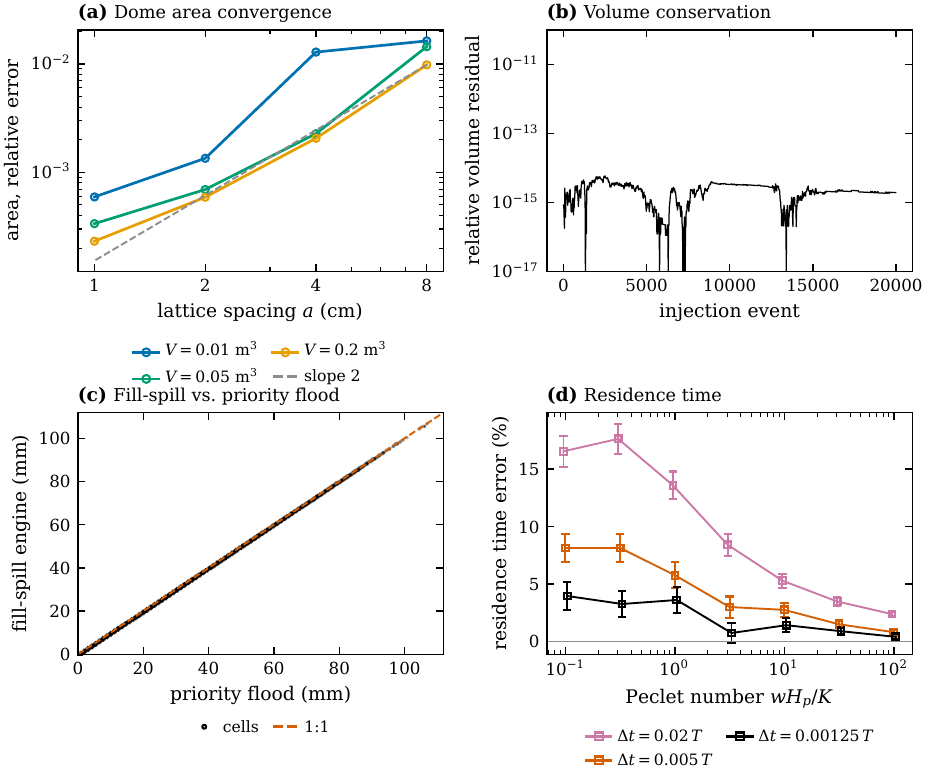}
    \caption{Verification. (a) Relative error of the covered area of a
    paraboloidal dome against lattice spacing for three gas volumes, with a
    slope-2 reference. (b) Relative volume residual after each of 20\,000
    random injections on a dipping rough ceiling. (c) Saturated gas
    thickness from the fill-spill engine against the priority-flood
    filling, all cells of twelve ceilings with dips from 0 to $8^{\circ}$.
    (d) Relative error of the Lagrangian mean residence time against the
    exact result of Eq.~\eqref{eq:residence}, for three time steps given
    as fractions of the exact residence time $T$; bars are two standard
    errors.}
    \label{fig:verification}
\end{figure}

The computation passes every verification test (Fig.~\ref{fig:verification},
Table~\ref{tab:verification}). The saturated state of the event-driven
computation agrees with the independent priority-flood filling of
Eq.~\eqref{eq:flood} with a maximum absolute difference of zero, over
147\,456 cells of which 27\,086 hold gas, on twelve ceilings with dips from
0 to $8^{\circ}$. The agreement is exact, not approximate, because both
algorithms fill every depression of $-h_{\rm eff}$ to the elevation of its
outlet, and the elevation of an outlet is a value taken from the same
lattice. Over 20\,000 random injections, the relative volume residual
remains below $6\times10^{-15}$, close to the precision of double-precision
arithmetic summed over many pools, which confirms that the bookkeeping of
Eqs.~\eqref{eq:volume} and \eqref{eq:dlevel} neither creates nor loses gas
at spills or merges. The covered area of the paraboloidal dome converges to
its exact value with an observed order between 1.3 and 2.3, reaching a
relative error of $2.3\times10^{-4}$ at $a = 1$~cm for the largest volume.
The order exceeds one because the contact line is a closed curve whose
pixelation errors partly cancel around the circumference. The Lagrangian
residence time exceeds Eq.~\eqref{eq:residence} by a positive bias. The bias
is largest at small P\'eclet number, where grains approach the floor many
times by diffusion before crossing it, and it halves when the step is
reduced by a factor of four, as expected for a bias of order
$(\Delta t)^{1/2}$ that arises from detecting floor crossings only at the
ends of steps. For the smallest step it is below 4 percent. In the
reference computations, the step is a small fraction of the residence time
of the fine classes that control visibility, and only the arrival of the
coarsest class at the floor may be late by a few percent.

\begin{table}[H]
\centering
\caption{Measured residuals and exponents. Intervals are 95 percent
bootstrap intervals.}
\label{tab:verification}
\small
\begin{tabular}{lll}
\toprule
Test & Expected & Measured \\
\midrule
Engine vs priority flood, max abs (m) & 0 & 0 \\
Volume residual, 20\,000 events & 0 & $5.9\times10^{-15}$ \\
Dome area, $a = 8$ to 1~cm, largest $V$ & $\to 0$ &
$9.8\times10^{-3}$ to $2.3\times10^{-4}$ \\
Cluster dimension $D_{f}$ & 1.8958 & 1.878 [1.847, 1.907] \\
Acceptance half-density & $p_{c} = 0.5927$ & 0.588 \\
Finger exponent & $-0.5714$ & $-0.578$ [$-0.595$, $-0.562$] \\
Relief exponent, $H = 0.5$ / 0.8 & 0.5 / 0.8 & 0.511 / 0.767 \\
Point source $A \propto V^{x}$, $H = 0.5$ & 0.800 & 0.784 [0.745, 0.820] \\
Point source $A \propto V^{x}$, $H = 0.8$ & 0.714 & 0.727 [0.691, 0.765] \\
Pools $V \propto A^{x}$, $H = 0.5$ & 1.250 & 1.304 [1.292, 1.316] \\
Pools $V \propto A^{x}$, $H = 0.8$ & 1.400 & 1.435 [1.416, 1.455] \\
Residence time, smallest step & 0 & $+0.4$ to $+4.0$ percent \\
\bottomrule
\end{tabular}
\end{table}

\begin{figure}[H]
    \centering
    \includegraphics[width=\linewidth]{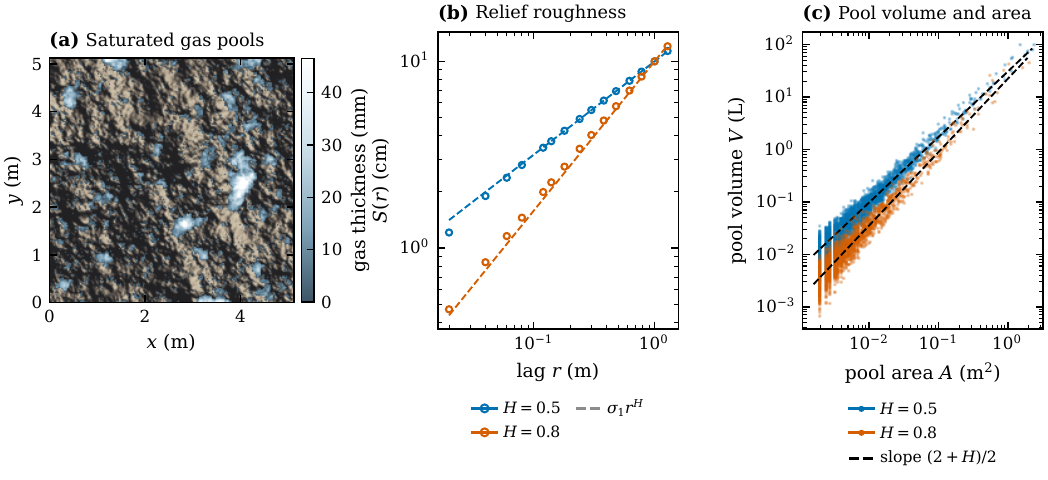}
    \caption{The ceiling and the pools it can hold. (a) Effective ceiling
    relief (hillshade) and gas thickness of the saturated state, in which
    every inverted depression is full to its spill point, for $H = 0.8$,
    $\sigma_{1} = 0.10$~m, and no dip. (b) Structure function $S(r)$ for
    $H = 0.5$ and 0.8, eight realizations each, with $\sigma_{1}r^{H}$
    (dashed). (c) Volume against area of every pool with at least five
    cells in the saturated states, with slope $(2+H)/2$ (dashed).}
    \label{fig:ceiling}
\end{figure}

The synthetic ceilings follow Eq.~\eqref{eq:structure}
(Fig.~\ref{fig:ceiling}b). The fitted exponent is 0.511 for an input of 0.5
and 0.767 for an input of 0.8. The second is reduced because the spectrum is
truncated at the lattice scale, which removes the steepest small-scale
relief and lowers the apparent exponent most when $H$ is close to one. When
every pool of an untilted ceiling is full, gas covers isolated pools of all
sizes up to that of the domain (Fig.~\ref{fig:ceiling}a). The largest pools
lie over the long-wavelength highs of the relief, and the smallest occupy
local highs only a few cells across. Across these pools, the volume grows
with the area as predicted by Eq.~\eqref{eq:areavolume}
(Fig.~\ref{fig:ceiling}c). The fitted exponents, 1.304 for $H = 0.5$ and
1.435 for $H = 0.8$ from 1939 and 1042 pools of at least ten cells, exceed
the closed-form values of 1.25 and 1.40 by 3 to 4 percent, and their
intervals exclude them. We attribute the difference to the reduced relief
exponent at the lattice scale, which affects the smallest pools most and
makes them shallower than the self-affine scaling implies. The slope of the
relation nevertheless separates the two Hurst exponents clearly, so the
relation between the volume and area of pools is a sensitive measure of the
roughness of the ceiling.

\begin{figure}[H]
    \centering
    \includegraphics[width=\linewidth]{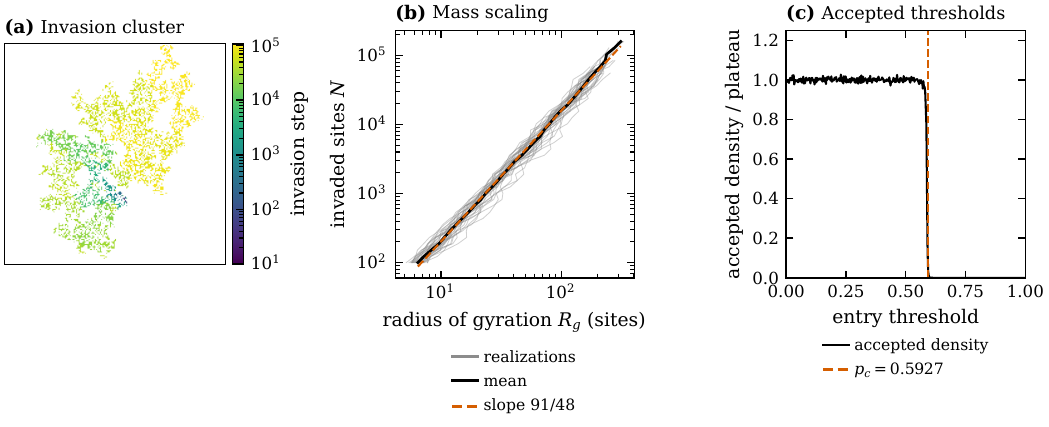}
    \caption{The connected capillary limit. (a) One invasion-percolation
    cluster on a $1201^{2}$ lattice, colored by invasion step. (b) Invaded
    sites against radius of gyration during growth, 32 realizations and
    their mean, with slope 91/48 (dashed). (c) Density of accepted
    thresholds normalized by its plateau, with $p_{c}$ (dashed).}
    \label{fig:ip}
\end{figure}

\begin{figure}[H]
    \centering
    \includegraphics[width=\linewidth]{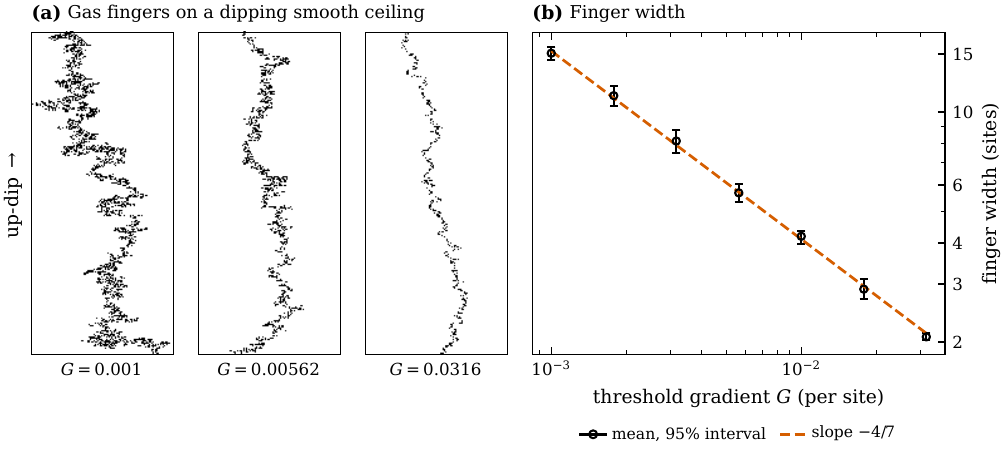}
    \caption{Gas on a dipping smooth ceiling. (a) Invasion fingers for
    three threshold gradients $G$, on strips 800 sites wide. (b) Root-mean-
    square lateral width of the finger against $G$, means of eight
    realizations with 95 percent bootstrap intervals, with slope $-4/7$
    (dashed).}
    \label{fig:gradient}
\end{figure}

In the connected capillary limit, the invaded set reproduces the exact
properties of critical percolation (Fig.~\ref{fig:ip}). The number of
invaded sites grows with the radius of gyration with an exponent of 1.878,
whose interval [1.847, 1.907] contains 91/48. The density of accepted
thresholds is uniform below $p_{c}$ and falls to half its plateau at 0.588,
slightly below $p_{c}$ because each run stops when the cluster first reaches
the edge of the lattice, before the invasion has accepted the full share of
thresholds just below $p_{c}$. The color sequence of
Fig.~\ref{fig:ip}a shows how the cluster grows. Long periods of slow,
local advance alternate with bursts in which the front runs through a chain
of easy sites, the lattice counterpart of the pore-scale jumps observed in
slow drainage \cite{Maloy1992}. On a dipping smooth ceiling, the gas rises
in a single finger whose width decreases with the threshold gradient
(Fig.~\ref{fig:gradient}). The measured exponent, $-0.578$ with an
interval of [$-0.595$, $-0.562$], contains $-4/7$, as found in experiments
and simulations of gravity-destabilized invasion
\cite{Birovljev1991, Meakin1992}. The fingers in Fig.~\ref{fig:gradient}a
show the internal structure of percolation clusters up to the finger width,
with loops and dead ends, and a nearly straight path above it. On a smooth
ceiling with even a weak dip, exhaled gas therefore follows a narrow track
up-dip, and the track becomes narrower as the dip or the density contrast
increases relative to the spread of pinning pressures.

\begin{figure}[H]
    \centering
    \includegraphics[width=\linewidth]{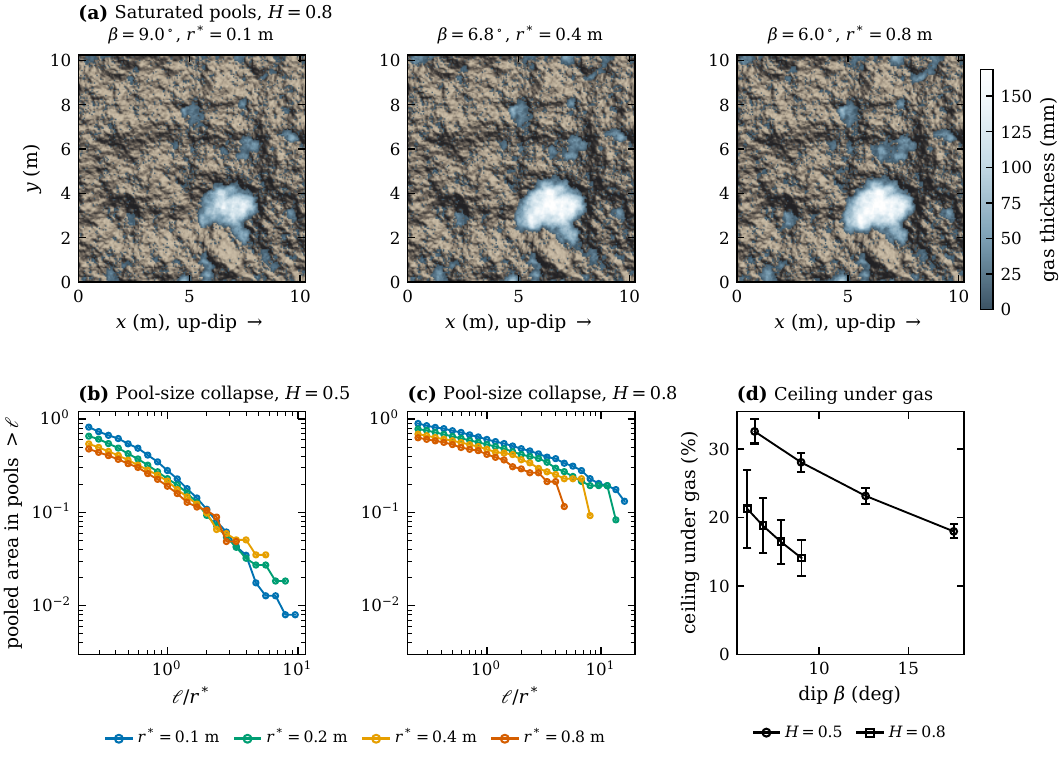}
    \caption{Pools on dipping rough ceilings, $512\times512$ cells
    (10.24~m square), $\sigma_{1} = 0.10$~m, no arch or pinning, six
    realizations per dip. (a) Saturated states for $H = 0.8$ at three dips.
    (b, c) Fraction of the pooled area held in pools of linear size
    $\sqrt{A}$ larger than $\ell$, against $\ell/r^{*}$, for $H = 0.5$ and
    0.8. (d) Fraction of the ceiling that holds gas against dip; bars are
    one standard deviation.}
    \label{fig:tilt}
\end{figure}

On rough dipping ceilings, the pool scale $r^{*}$ organizes the pool sizes
but does not cap them (Fig.~\ref{fig:tilt}). For $H = 0.5$, the fraction of
pooled area held in pools larger than $\ell$ collapses onto a single curve of
$\ell/r^{*}$ above about $1.5\,r^{*}$, for values of $r^{*}$ from 0.1 to
0.8~m that correspond to dips from $17.5^{\circ}$ to $6.4^{\circ}$. Below
$1.5\,r^{*}$, the curves separate because pools only a few cells across are
affected by the lattice and the pinning noise. For $H = 0.8$, the curves do
not collapse within a domain of 10~m. The stretched exponential of
Eq.~\eqref{eq:soft} decays with an exponent of 0.4 in this case, so pools of
ten times $r^{*}$ remain likely and the largest pools are limited by the
size of the domain rather than by $r^{*}$. The same large pool persists in
Fig.~\ref{fig:tilt}a as the dip increases from $6.0^{\circ}$ to
$9.0^{\circ}$, shrinking at its margins but not disappearing, because it
occupies a long-wavelength high whose relief exceeds the rise of the dip
across it. The fraction of the ceiling that can hold gas decreases with the
dip (Fig.~\ref{fig:tilt}d), from 32.6 to 18.0 percent between $6.4^{\circ}$
and $17.5^{\circ}$ for $H = 0.5$, and from 21.3 to 14.1 percent between
$6.0^{\circ}$ and $9.0^{\circ}$ for $H = 0.8$. The rougher ceiling at small
scale, $H = 0.5$, holds more gas at a given dip because its relief reverses
more often over short distances. Because $r^{*}$ varies with the fifth power
of $\sigma_{1}/(r_{0}\tan\beta)$ for $H = 0.8$, a change of a few degrees in
dip changes the pool scale by orders of magnitude, from 25~m at
$3^{\circ}$ to 0.18~m at $8^{\circ}$ for $\sigma_{1} = 0.10$~m.

\begin{figure}[H]
    \centering
    \includegraphics[width=\linewidth]{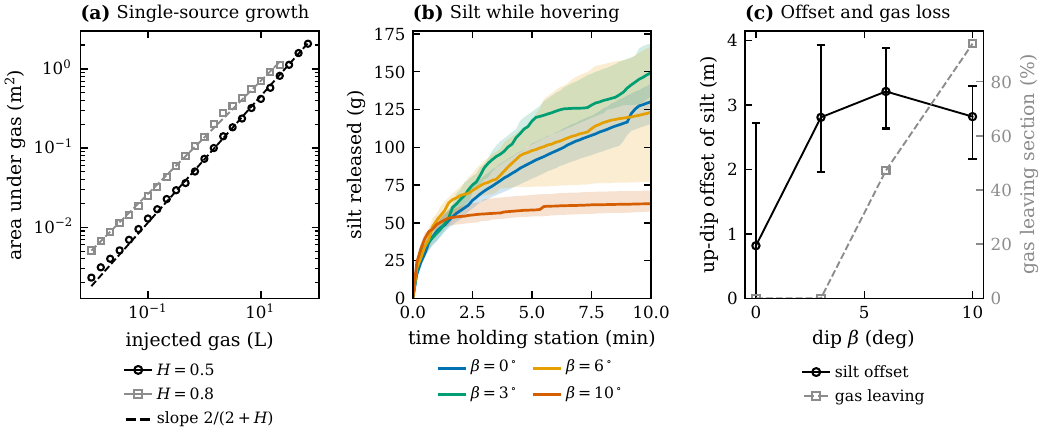}
    \caption{Gas released at one place. (a) Covered area against injected
    volume for a point source on ceilings with no dip, arch, or pinning,
    eight realizations per Hurst exponent, with slope $2/(2+H)$ (dashed).
    (b) Silt released within the 12.8~m section by one diver
    (RMV 20~L\,min$^{-1}$) holding station for 10~min in the reference
    passage at four dips, median and interquartile range of twelve
    ceilings. (c) Mass-weighted up-dip distance of the released silt from
    the diver (left axis) and fraction of the exhaled gas leaving the
    section (right axis).}
    \label{fig:hover}
\end{figure}

Gas released at one point covers an area that grows as $V^{0.784}$ for
$H = 0.5$ and $V^{0.727}$ for $H = 0.8$ (Fig.~\ref{fig:hover}a). Both
intervals contain the exponents of Eq.~\eqref{eq:areavolume}, so the second
relation holds even though the first deviates slightly, which suggests that
the growth of a single cluster is dominated by its larger pools, where the
lattice cutoff matters least. A diver who holds station therefore covers
progressively less new ceiling with each breath, and the silt released
slows with time (Fig.~\ref{fig:hover}b). Within the first minute the release
is nearly the same at every dip, because the first breaths fill the pools
directly above the diver. Thereafter the curves separate. After 10~min, the
median release within the modeled section is 130~g on a level ceiling,
149~g at $3^{\circ}$, 123~g at $6^{\circ}$, and 63~g at $10^{\circ}$. A
small dip increases the release because gas leaving the first pools travels
along the ceiling and strips new paths. A larger dip reduces the release
within the section because the gas escapes from it, 47 percent at
$6^{\circ}$ and 94 percent at $10^{\circ}$, and continues to strip silt
farther up-dip. The silt does not fall where the diver hovers
(Fig.~\ref{fig:hover}c). Its mass-weighted position lies 0.8~m up-dip of the
diver on a level ceiling, where the offset arises only from the relief and
the arch, and 2.8 to 3.2~m up-dip at dips of 3 to $10^{\circ}$, the last
value being limited by the end of the section. The spread among ceilings is
large, with interquartile ranges of about one meter, because the position of
the nearest closed high up-dip of the diver differs from one ceiling to the
next.

\begin{figure}[H]
    \centering
    \includegraphics[width=\linewidth]{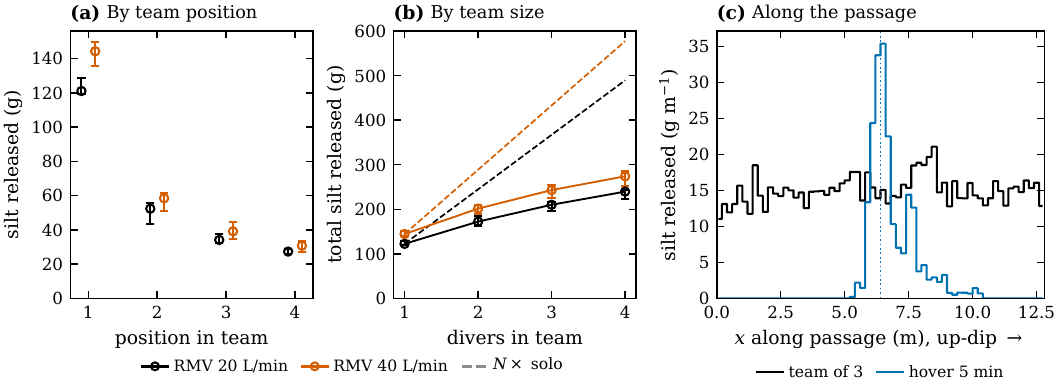}
    \caption{Silt released by a team. Reference passage, twelve ceilings,
    divers swimming up-dip at 0.2~m\,s$^{-1}$, 20~s apart. (a) Silt
    released by each diver of a team of four, for RMV 20 and
    40~L\,min$^{-1}$, median and interquartile range. (b) Total silt
    released by teams of one to four divers, with $N$ times the solo median
    (dashed). (c) Silt released per meter of passage by a team of three and
    by one diver holding station for 5~min at $x = 6.4$~m, medians.}
    \label{fig:team}
\end{figure}

When a team swims through the reference passage, the silt released by each
diver depends strongly on position (Fig.~\ref{fig:team}a). The four divers
of a team release 121, 52, 34, and 27~g in order, so that the lead diver
releases 2.3 times as much as the second, and each following diver releases
less than the one before. Doubling the respiratory minute volume to
40~L\,min$^{-1}$ increases the release of the lead diver by 19 percent, to
144~g, and of the second diver by 11 percent, and does not change the
ordering. The total released by the team grows sublinearly with its size
(Fig.~\ref{fig:team}b), from 122~g for one diver to 172, 210, and 240~g for
two, three, and four divers, a ratio of 1.96 between four divers and one.
The mechanism is visible in the gas states. The gas of the lead diver fills
new pools and opens new migration paths, and the contact lines it moves
cross clean rock for the first time. The gas of the following divers arrives
at the same places, because they follow the same line, and largely enters
pools that are already full and travels paths that have already been
stripped. Their contribution comes mainly from the margins of the existing
pools and from new paths opened when the added gas makes a pool spill
elsewhere. Along the passage, a team in transit releases silt almost
uniformly, at about 15~g per meter of passage, whereas a diver who holds
station for 5~min concentrates a release of 99~g within about 2~m of the
hovering position, with a peak of about 35~g per meter located up-dip of
the diver (Fig.~\ref{fig:team}c). The hovering diver thus releases less
silt in total than a team of three in transit, 99~g against 210~g, but
deposits it in a section one-sixth as long.

The ordering by position survives every combination of detachment constants
and ceiling geometry that we tested (Table~\ref{tab:sensitivity}). Over
contact-line fractions $\phi_{c}$ from 0.2 to 0.8 and stripping volumes
$v_{d}$ from 0.2 to 20~mL, the lead diver releases 2.3 to 2.5 times the silt
of the second, and a team of four releases 1.86 to 1.92 times the silt of a
solo diver. The absolute mass scales almost in proportion to $\phi_{c}$,
while $v_{d}$ has little effect because a single breath of 1.37~L exceeds
even the largest value by a factor of about seventy, so that any path
crossed by a breath is stripped almost completely. Varying the relief
amplitude from 0.05 to 0.20~m, the pinning radius from 2 to 8~mm, removing
the arch, or increasing the dip to $6^{\circ}$ changes the ratio of lead to
second diver to between 2.2 and 2.8 and the ratio of team to solo to between
1.79 and 1.97. The ratio of lead to second diver is largest on the smoothest
ceiling, $\sigma_{1} = 0.05$~m, where pools are small and the lead diver's gas
reaches and strips a larger fraction of the accessible paths, and it is
insensitive to the pinning radius, which changes the effective ceiling by
millimeters against a relief of centimeters.

\begin{table}[H]
\centering
\caption{Sensitivity of the team results, team of four, medians of six
ceilings. Ratios are lead to second diver and team of four to solo
diver.}
\label{tab:sensitivity}
\small
\begin{tabular}{lccc}
\toprule
Case & Lead diver (g) & Lead/second & Team/solo \\
\midrule
Reference ($\phi_{c} = 0.5$, $v_{d} = 2$~mL) & 120.7 & 2.36 & 1.90 \\
$\phi_{c} = 0.2$, $v_{d} = 2$~mL & 96.6 & 2.49 & 1.86 \\
$\phi_{c} = 0.8$, $v_{d} = 2$~mL & 148.1 & 2.35 & 1.91 \\
$\phi_{c} = 0.5$, $v_{d} = 20$~mL & 112.9 & 2.32 & 1.91 \\
$\sigma_{1} = 0.05$~m & 127.2 & 2.80 & 1.79 \\
$\sigma_{1} = 0.20$~m & 110.5 & 2.23 & 1.97 \\
$r_{50} = 2$~mm & 120.6 & 2.36 & 1.90 \\
$r_{50} = 8$~mm & 121.6 & 2.36 & 1.89 \\
No arch & 137.4 & 2.40 & 1.91 \\
Dip $6^{\circ}$ & 120.9 & 2.48 & 1.88 \\
\bottomrule
\end{tabular}
\end{table}

\begin{figure}[H]
    \centering
    \includegraphics[width=\linewidth]{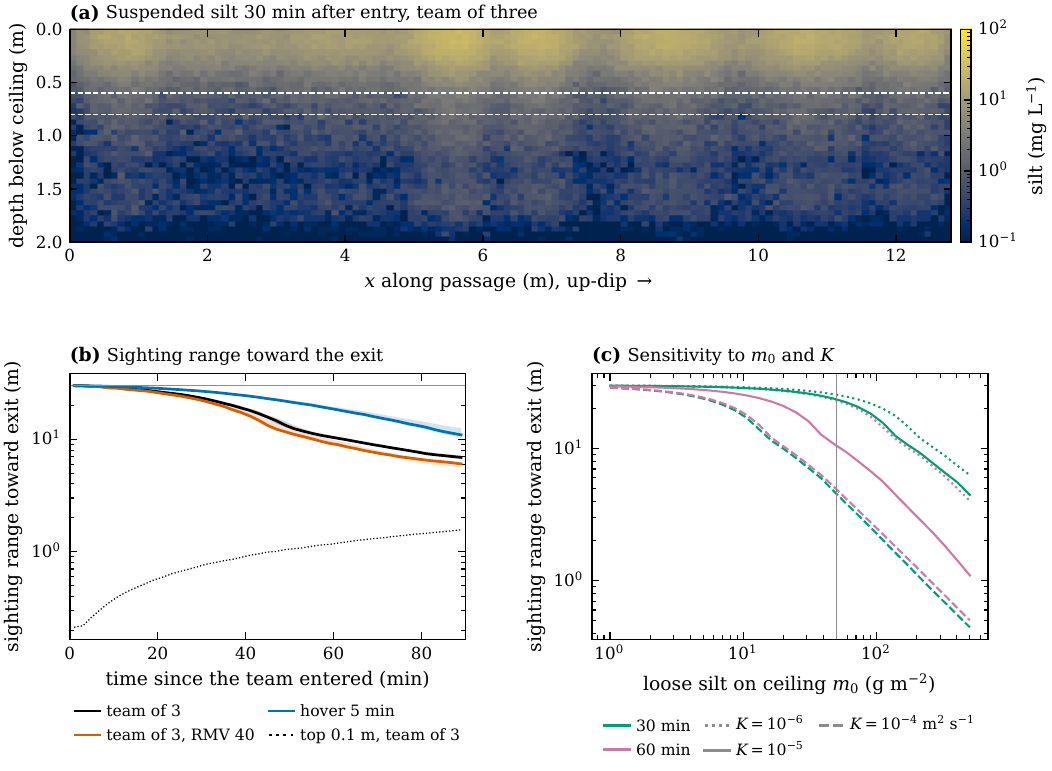}
    \caption{Visibility after the team has passed. Released silt settles as
    22 Stokes classes with $K = 10^{-5}$~m$^{2}$\,s$^{-1}$ in the
    width-averaged reference passage. (a) Suspended silt 30~min after a
    team of three entered, one ceiling, logarithmic scale; dashed lines
    bound the eye band. (b) Smallest sighting range toward the exit along
    the passage in the eye band, median and interquartile range of six
    ceilings, for three scenarios, and in the top 0.1~m below the ceiling
    for the team of three (dotted). (c) Smallest eye-band sighting range 30
    and 60~min after entry against the loose silt inventory $m_{0}$, for
    three eddy diffusivities; the vertical line marks the reference
    inventory.}
    \label{fig:visibility}
\end{figure}

Silt released by a team of three spreads through the passage over the
following hour (Fig.~\ref{fig:visibility}). Coarse grains, above about
60~$\mu$m, settle the full 2~m height in about 10~min or less, whereas the fine fraction remains
in a layer under the ceiling and descends slowly. Thirty minutes after the
team entered, the concentration exceeds 10~mg\,L$^{-1}$ in the upper 0.3~m
and is of order 1~mg\,L$^{-1}$ at eye level (Fig.~\ref{fig:visibility}a).
The layer is not uniform along the passage. It is thickest beneath the
places where the gas pooled and spilled, and these maxima persist because
horizontal mixing over 30~min spreads silt by only a few tens of
centimeters at the reference diffusivity. The sighting range in the top
0.1~m below the ceiling stays below 1~m for about 45~min. At eye level, the
median sighting range toward the exit, taken at the worst point along the
passage, decreases from 27.7~m after 15~min to 23.6~m after 29~min,
15.7~m after 45~min, 10.5~m after 59~min, and 6.9~m after 89~min
(Fig.~\ref{fig:visibility}b). The decline accelerates between about 30 and
50~min, when grains of about 20 to 30~$\mu$m, whose Stokes fall time over the 1.3~m from
ceiling to eye level is 25 to 60~min, which carry a large share of the
mass and attenuate efficiently, reach eye level. For a team breathing
40~L\,min$^{-1}$, the corresponding values are 27.3, 22.6, 13.3, 9.2, and
6.0~m. After a diver holds station for 5~min, the decrease is slower, to
18.9~m after 59~min and 10.9~m after 89~min, because less silt is released
and it is confined to a shorter part of the passage, which a sight line
toward the exit crosses for a shorter distance. In every case, visibility at
eye level continues to worsen for an hour or more after the divers have
passed, which is the period during which a team would typically return
along the same passage.

The sighting ranges depend strongly on two quantities that are not known for
caves (Fig.~\ref{fig:visibility}c). Because the released mass is
proportional to $m_{0}$, the attenuation added by silt scales with it, and
at the reference diffusivity the eye-band sighting range after 60~min
decreases from 25.5~m for $m_{0} = 10$~g\,m$^{-2}$ to 10.6~m at the reference
inventory and 3.1~m at 200~g\,m$^{-2}$. The dependence on the eddy
diffusivity is equally strong but less intuitive. At the reference
inventory, the same quantity is 23.0~m for $K = 10^{-6}$~m$^{2}$\,s$^{-1}$
and 4.9~m for $K = 10^{-4}$~m$^{2}$\,s$^{-1}$. Stronger mixing carries the fine
fraction away from the ceiling and down to eye level sooner, and because the
finest grains settle more slowly than they mix, it then keeps them
distributed through the water column instead of in a layer that the eye
band lies beneath. In nearly still water the silt remains concentrated near
the ceiling, which is opaque but lies above the eyes of a diver in a passage
of this height.

\section{Discussion}
\label{sec:discussion}

The results separate two kinds of control on percolation, geometric and
operational. The geometric control is the balance between the relief of the
ceiling and its dip, which Eq.~\eqref{eq:rstar} condenses into one length.
On rough ceilings with little dip, gas pools near where it arrives, and the
silt it releases falls close to the diver. On steeper ceilings, gas leaves
the first pools along migration paths and strips silt meters away from its
source, in places that a diver may not associate with the bubbles. Because
$r^{*}$ depends on the dip through the power $1/(1-H)$, which is large when
the relief is persistent, the transition between these behaviors can occur
over a few degrees. The cutoff is soft, however, and long-wavelength highs
continue to hold large pools on ceilings steeper than $r^{*}$ alone would
suggest. In this respect the problem resembles the gravity-modified invasion
of porous media, in which a hydrostatic gradient competes with a random
distribution of thresholds \cite{Wilkinson1984}. The difference is that the
thresholds here are themselves correlated over all scales by the
self-affine relief, so that the competition is decided scale by scale and
yields a power law in the dip rather than a single crossover.

The ceiling also places the problem between two limits of immiscible
displacement. Where the relief of a single cell exceeds the pinning height,
gas behaves as a gravity-dominated fluid that fills and spills, and its
pools fragment and reconnect in the manner described for buoyancy-driven
migration \cite{Wagner1997}. Where the ceiling is smooth, pinning dominates,
gas remains connected, and the advance is invasion percolation in the
capillary-fingering regime of the displacement phase diagram
\cite{Lenormand1988}, with a finger width set by the dip
\cite{Birovljev1991, Meakin1992}. The ratio of relief at the lattice scale to
the pinning height, close to one in the reference case, marks the boundary.
Real ceilings may span both regimes, with smooth flowstone or bedding planes
adjacent to scalloped or fractured rock. The model treats the two limits
separately and does not represent the crossover between them, which would
require resolving the meniscus at the scale of the pinning height.

The operational control is the history of the ceiling. The amount of silt
released depends far more on whether a ceiling has already been swept than
on how much gas reaches it. This is a consequence of the hierarchy of the
fill-spill-merge process. The pools and paths available to gas are fixed by
the geometry, and the first gas to arrive occupies them. Later gas follows
the same routes, because the steepest-ascent paths and spill points do not
change, and it crosses rock that has already been stripped. Breathing harder
adds gas mainly to pools and paths that already exist, which explains why
doubling the respiratory minute volume increases the release of the lead
diver by only 19 percent and why the release from a diver who holds station
slows with time in the manner set by Eq.~\eqref{eq:areavolume}. The same
reasoning implies that the release from a passage is limited by its loose
inventory and its accessible area, not by the volume of gas, once the
accessible area has been swept.

The timing of the visibility loss follows from the physics of settling and
attenuation. The silt that controls visibility is the fine fraction, both
because fine grains attenuate light most per unit mass, as
Eq.~\eqref{eq:attenuation} shows and as observed in natural waters
\cite{Boss2009, DaviesColley2001}, and because they remain in suspension
longest, as Eq.~\eqref{eq:residence} shows. A team that passes beneath a
silty ceiling therefore leaves behind a layer that is opaque near the rock
and that descends toward eye level over the following tens of minutes.
Visibility on the return may then be worse than on entry even if the passage
appeared clear immediately behind the team. The strong dependence on the
eddy diffusivity shows that the residual mixing of a passage, including
the wake of the divers themselves, matters as much as the amount of silt.

These results are consistent with common practice in cave diving, although
they do not test it. Practice discourages lingering under silty ceilings,
and in the model a stationary diver concentrates the release in a short
section of the passage up-dip of the hovering position. Closed-circuit
rebreathers are valued in silty systems, and in the model a diver who
exhales no gas releases no silt from the ceiling. The model adds three
features that are less obvious from experience. First, the position of a
diver in a team affects the silt that diver releases more than the rate at
which the diver breathes. Second, on a dipping passage the silt released by
a hovering diver falls meters up-dip, so its effect may be met by a team
continuing in that direction rather than by the diver who caused it.
Third, visibility at eye level continues to decline long after the team has
passed. We emphasize that these are properties of the model under the
assumptions stated here, and that their relevance to real caves depends on
measurements that have not yet been made.

The model has several limitations. The gas is quasi-static. Bubble dynamics,
inertia at spill points, the dissolution of trapped gas, and receding
contact lines are not represented, and the volume of the pinned meniscus is
neglected. Contact-angle hysteresis is represented only through a pinning
height for advancing lines \cite{JoannyDeGennes1984}. Pools are
disconnected fill-spill pools, the connected capillary limit is computed
separately, and the crossover between the two is not modeled. Silt detached
under a pool is released to the water at once, although some may remain on
the interface or fall onto the rock beneath the pool. The detachment
efficiencies are not derived from particle-scale forces but represent the
outcome of competing capillary and adhesive forces whose magnitude depends
on particle shape and surface chemistry \cite{Sharma2008, Aramrak2013}. The
water column is averaged across the width of the passage, which reduces
local peaks of concentration and therefore overestimates the sighting range
near the centerline. The eddy diffusivity is constant, whereas mixing in a
passage after a team has passed is likely to decay with time, a situation
for which random-walk models require additional care \cite{Visser1997}.
Settling follows the Stokes law, which overestimates the velocity of the
coarsest class \cite{FergusonChurch2004}, and flocculation, which can
increase the settling velocity of fine particles by orders of magnitude in
saline or organic-rich water \cite{Winterwerp2002}, is not included. Floc
formation would shorten the persistence of the fine layer in anchialine or
marine caves and may make the reference results conservative there. Finally,
the loose silt inventory, the detachment constants, the pinning radius, the
statistics of the relief, the grain sizes, and the eddy diffusivity are
illustrative, and no measurement of them on a cave ceiling is known to us.
The absolute silt masses scale with $m_{0}$, and the sighting ranges change
by an order of magnitude between $K = 10^{-6}$ and
$10^{-4}$~m$^{2}$\,s$^{-1}$.

Several of the results are, in contrast, independent of these choices. The
agreement of the event-driven computation with priority flooding, the
conservation of volume, the percolation limits, the pool scale, and the
area-volume exponents are properties of the geometry. The ordering of divers
by position held for every combination of detachment constants and ceiling
geometry tested, because it follows from the reuse of a fixed hierarchy of
pools and paths rather than from the values of the constants. These
findings suggest measurements that would test the model. Laboratory
experiments with gas injected beneath inverted rough plates of known
self-affine relief could test the pool scale and the area-volume law
directly. Repeated photographs of a silty cave ceiling before and after the
passage of divers could measure stripping efficiencies. Transmissometer
profiles taken at fixed points in a passage after a team has passed could
test the predicted timing of visibility loss and constrain the eddy
diffusivity.

\section{Conclusions}
\label{sec:conclusion}

We developed an idealized model of percolation, the release of silt by
exhaled gas beneath the ceilings of submerged caves. The model combines an
exact fill-spill-merge description of buoyant gas on a self-affine rough
ceiling with capillary pinning of the contact line, detachment of silt by
moving contact lines, Lagrangian settling of a spectrum of grain sizes, and
the beam-attenuation criterion for the sighting range of a diver. The
computation agrees exactly with an independent depression-filling algorithm,
conserves gas volume to round-off, and recovers the known exponents of
invasion percolation and of gravity-destabilized invasion in its capillary
limit. A balance between relief and dip defines a pool scale that organizes
where gas collects, and the area that gas from a single source covers grows
more slowly than the volume supplied, in agreement with the self-affine
geometry. The cutoff at the pool scale is soft, so large pools persist on
ceilings steeper than the scale alone suggests. The silt released by a team
depends mainly on the history of the ceiling. The lead diver releases far
more than the diver behind, later divers add progressively
less, and breathing harder adds comparatively little, because the gas of
later divers reuses pools and paths that the lead diver has already filled
and stripped. A diver who holds station concentrates the release in a short
section up-dip of the hovering position. Visibility at eye level toward the
exit continues to decline for an hour or more after a team has passed, as
fine silt descends from a nearly opaque layer beneath the ceiling. These
conclusions hold across the detachment efficiencies and ceiling geometries
tested, whereas absolute silt masses and sighting ranges depend on the loose
silt inventory and on mixing in still cave water. Measurements of the loose
silt on cave ceilings, of the efficiency with which moving contact lines
remove it, and of mixing after the passage of divers would allow these
comparisons to be converted into predictions for particular passages.

\section*{Declaration of competing interest}
The authors declare that they have no known competing financial interests or
personal relationships that could have appeared to influence the work
reported in this article.

\section*{Declaration of generative AI use}
During the preparation of this work, the authors used Claude Sonnet 5 (Anthropic) solely to refine the grammar and vocabulary of the manuscript. The authors reviewed and edited the resulting text and take full responsibility for the conceptualization, model development, mathematical derivations, computational results, and their interpretation.

\section*{Code and data availability}
This study uses no external data. The analysis code and all of its outputs,
including CSV tables of the data behind each figure panel, the figures in
vector and raster form, plain-text computation reports, animations, and the
scripts that regenerate them, are openly available at
\url{https://github.com/sandyherho/cave_percolation} under the MIT License.

\section*{Funding}
This study was supported by the Faculty of Earth Sciences and Technology
(FITB), Bandung Institute of Technology (ITB), through the PPMI Research
Program 2026 under Project ID FITB.PPMI-1-19-2026.


\begin{thebibliography}{99}

\bibitem{Aramrak2013} Aramrak, S.; Flury, M.; Harsh, J. B.; Zollars, R. L.;
Davis, H. P. Does Colloid Shape Affect Detachment of Colloids by a Moving Air–Water Interface? \textit{Langmuir} \textbf{2013}, \textit{29(19)},
5770--5780. \url{https://doi.org/10.1021/la400252q}

\bibitem{Barnes2021} Barnes, R.; Callaghan, K. L.; Wickert, A. D. Computing
water flow through complex landscapes, Part 3: Fill-Spill-Merge: flow
routing in depression hierarchies. \textit{Earth Surf. Dynam.}
\textbf{2021}, \textit{9(1)}, 105.
\url{https://doi.org/10.5194/esurf-9-105-2021}

\bibitem{Birovljev1991} Birovljev, A.; Furuberg, L.; Feder, J.;
J{\o}ssang, T.; Mly, K. J.; Aharony, A. Gravity invasion
percolation in two dimensions: Experiment and simulation. \textit{Phys.
Rev. Lett.} \textbf{1991}, \textit{67(5)}, 584--587.
\url{https://doi.org/10.1103/PhysRevLett.67.584}

\bibitem{Boss2009} Boss, E.; Taylor, L.; Gilbert, S.; Gundersen, K.;
Hawley, N.; Janzen, C.; Johengen, T.; Purcell, H.; Robertson, C.;
Schar, D. W. H.; Smith, G. J.; Tamburri, M. N. Comparison of inherent
optical properties as a surrogate for particulate matter concentration in
coastal waters. \textit{Limnol. Oceanogr. Methods} \textbf{2009},
\textit{7}, 803--810. \url{https://doi.org/10.4319/lom.2009.7.803}

\bibitem{Buzzacott2009} Buzzacott, P. L.; Zeigler, E.; Denoble, P.;
Vann, R. American Cave Diving Fatalities 1969-2007. \textit{Int. J.
Aquat. Res. Educ.} \textbf{2009}, \textit{3(2)}, 162--177.
\url{https://doi.org/10.25035/ijare.03.02.07}

\bibitem{Candela2012} Candela, T.; Renard, F.; Klinger, Y.; Mair, K.;
Schmittbuhl, J.; Brodsky, E. E. Roughness of fault surfaces over nine
decades of length scales. \textit{J. Geophys. Res. Solid Earth}
\textbf{2012}, \textit{117}, B08409.
\url{https://doi.org/10.1029/2011JB009041}

\bibitem{DaviesColley2001} Davies-Colley, R. J.; Smith, D. G. TURBIDITY SUSPENDED SEDIMENT, AND WATER CLARITY: A REVIEW. \textit{J. Am. Water
Resour. Assoc.} \textbf{2001}, \textit{37(5)}, 1085--1101.
\url{https://doi.org/10.1111/j.1752-1688.2001.tb03624.x}

\bibitem{deGennes1985} de Gennes, P. G. Wetting: statics and dynamics.
\textit{Rev. Mod. Phys.} \textbf{1985}, \textit{57(3)}, 827--863.
\url{https://doi.org/10.1103/RevModPhys.57.827}

\bibitem{FergusonChurch2004} Ferguson, R. I.; Church, M. A Simple Universal Equation for Grain Settling Velocity. \textit{J. Sediment. Res.}
\textbf{2004}, \textit{74(6)}, 933--937.
\url{https://doi.org/10.1306/051204740933}

\bibitem{FordWilliams2007} Ford, D.; Williams, P. \textit{Karst Hydrogeology
and Geomorphology}; Wiley, \textbf{2007}.
\url{https://doi.org/10.1002/9781118684986}

\bibitem{Harris2020} Harris, C. R.; Millman, K. J.; van der Walt, S. J.;
Gommers, R.; Virtanen, P.; Cournapeau, D.; Wieser, E.; Taylor, J.;
Berg, S.; Smith, N. J.; Kern, R.; Picus, M.; Hoyer, S.; van Kerkwijk, M. H.;
Brett, M.; Haldane, A.; del R{\'\i}o, J. F.; Wiebe, M.; Peterson, P.;
G{\'e}rard-Marchant, P.; Sheppard, K.; Reddy, T.; Weckesser, W.;
Abbasi, H.; Gohlke, C.; Oliphant, T. E. Array programming with NumPy.
\textit{Nature} \textbf{2020}, \textit{585}, 357--362.
\url{https://doi.org/10.1038/s41586-020-2649-2}

\bibitem{Herho2026wave} Herho, S. H. S.; Anwar, I. P.; Khadami, F.; Ndruru,
T. R. E. B. N.; Suwarman, R.; Irawan, D. E. WAVE-ATTENUATION-1D: AN IDEALIZED ONE-DIMENSIONAL
FRAMEWORK FOR WAVE ATTENUATION THROUGH COASTAL
VEGETATION USING NUMBA-ACCELERATED SHALLOW
WATER EQUATIONS. \textit{J. Theor. Appl.
Mech.} \textbf{2026}, \textit{56}, 89--102.
\url{https://doi.org/10.55787/jtams.2026.1.AI00236}

\bibitem{Herho2025kh2d} Herho, S. H. S.; Trilaksono, N. J.; Fajary, F. R.;
Napitupulu, G.; Anwar, I. P.; Khadami, F.; Irawan, D. E. kh2d-solver: A
Python library for idealized two-dimensional incompressible Kelvin-Helmholtz
instability. \textit{Appl. Comput. Mech.} \textbf{2025}, \textit{19(2)},
125--156.
\url{https://doi.org/10.24132/acm.2025.1040}

\bibitem{Hunter2007} Hunter, J. D. Matplotlib: A 2D graphics environment.
\textit{Comput. Sci. Eng.} \textbf{2007}, \textit{9(3)}, 90--95.
\url{https://doi.org/10.1109/MCSE.2007.55}

\bibitem{Irawan2026amerta} Irawan, D. E.; Herho, S. H. S.; Anwar, I. P.;
Khadami, F.; Pamumpuni, A.; Kartiko, R. D.; Riawan, E.; Suwarman, R.;
Puradimaja, D. J. amerta: A Python library for idealized 1D Saint-Venant
dam-break simulation. \textit{Front. Water} \textbf{2026}, \textit{8},
1900409.
\url{https://doi.org/10.3389/frwa.2026.1900409}

\bibitem{Irawan2026kdv} Irawan, D. E.; Herho, S. H. S.; Pamumpuni, A.;
Kartiko, R. D.; Khadami, F.; Anwar, I. P.; Sujatmiko, K. A.; Handayani,
A. P.; Fajary, F. R.; Suwarman, R. An Open-Source Pseudo-Spectral Solver for
Idealized Korteweg--de Vries Soliton Simulations. \textit{Water}
\textbf{2026}, \textit{18(7)}, 779.
\url{https://doi.org/10.3390/w18070779}

\bibitem{JoannyDeGennes1984} Joanny, J. F.; de Gennes, P. G. A model for
contact angle hysteresis. \textit{J. Chem. Phys.} \textbf{1984},
\textit{81(1)}, 552--562. \url{https://doi.org/10.1063/1.447337}

\bibitem{Lenormand1988} Lenormand, R.; Touboul, E.; Zarcone, C. Numerical
models and experiments on immiscible displacements in porous media.
\textit{J. Fluid Mech.} \textbf{1988}, \textit{189}, 165--187.
\url{https://doi.org/10.1017/S0022112088000953}

\bibitem{Maloy1992} M{\aa}l{\o}y, K. J.; Furuberg, L.; Feder, J.;
J{\o}ssang, T. Dynamics of slow drainage in porous media. \textit{Phys.
Rev. Lett.} \textbf{1992}, \textit{68(14)}, 2161--2164.
\url{https://doi.org/10.1103/PhysRevLett.68.2161}

\bibitem{Meakin1992} Meakin, P.; Feder, J.; Frette, V.; J{\o}ssang, T.
Invasion percolation in a destabilizing gradient. \textit{Phys. Rev. A}
\textbf{1992}, \textit{46}, 3357.
\url{https://doi.org/10.1103/PhysRevA.46.3357}

\bibitem{NewmanZiff2000} Newman, M. E. J.; Ziff, R. M. Efficient Monte Carlo Algorithm and High-Precision Results for Percolation. \textit{Phys.
Rev. Lett.} \textbf{2000}, \textit{85(19)}, 4104--4107.
\url{https://doi.org/10.1103/PhysRevLett.85.4104}

\bibitem{Sharma2008} Sharma, P.; Flury, M.; Zhou, J. Detachment of colloids
from a solid surface by a moving air-water interface. \textit{J. Colloid
Interface Sci.} \textbf{2008}, \textit{326(1)}, 143--150.
\url{https://doi.org/10.1016/j.jcis.2008.07.030}

\bibitem{SpenceWoo2003} Spence, C.; Woo, M.-k. Hydrology of subarctic Canadian
shield: soil-filled valleys. \textit{J. Hydrol.} \textbf{2003},
\textit{279(1--4)}, 151--166.
\url{https://doi.org/10.1016/S0022-1694(03)00175-6}

\bibitem{Virtanen2020} Virtanen, P.; Gommers, R.; Oliphant, T. E.;
Haberland, M.; Reddy, T.; Cournapeau, D.; Burovski, E.; Peterson, P.;
Weckesser, W.; Bright, J.; van der Walt, S. J.; Brett, M.; Wilson, J.;
Millman, K. J.; Mayorov, N.; Nelson, A. R. J.; Jones, E.; Kern, R.;
Larson, E.; Carey, C. J.; Polat, {\.I}.; Feng, Y.; Moore, E. W.;
VanderPlas, J.; Laxalde, D.; Perktold, J.; Cimrman, R.; Henriksen, I.;
Quintero, E. A.; Harris, C. R.; Archibald, A. M.; Ribeiro, A. H.;
Pedregosa, F.; van Mulbregt, P.; SciPy 1.0 Contributors. SciPy 1.0:
fundamental algorithms for scientific computing in Python.
\textit{Nat. Methods} \textbf{2020}, \textit{17}, 261--272.
\url{https://doi.org/10.1038/s41592-019-0686-2}

\bibitem{Visser1997} Visser, A. W. Using random walk models to simulate the
vertical distribution of particles in a turbulent water column.
\textit{Mar. Ecol. Prog. Ser.} \textbf{1997}, \textit{158}, 275--281.
\url{https://doi.org/10.3354/meps158275}

\bibitem{Wagner1997} Wagner, G.; Meakin, P.; Feder, J.; J{\o}ssang, T.
Buoyancy-driven invasion percolation with migration and fragmentation.
\textit{Physica A} \textbf{1997}, \textit{245(3--4)}, 217--230.
\url{https://doi.org/10.1016/S0378-4371(97)00324-5}

\bibitem{Wilkinson1984} Wilkinson, D. Percolation model of immiscible
displacement in the presence of buoyancy forces. \textit{Phys. Rev. A}
\textbf{1984}, \textit{30(1)}, 520--531.
\url{https://doi.org/10.1103/PhysRevA.30.520}

\bibitem{WilkinsonWillemsen1983} Wilkinson, D.; Willemsen, J. F. Invasion
percolation: a new form of percolation theory. \textit{J. Phys. A: Math.
Gen.} \textbf{1983}, \textit{16}, 3365.
\url{https://doi.org/10.1088/0305-4470/16/14/028}

\bibitem{Winterwerp2002} Winterwerp, J. C. On the flocculation and settling
velocity of estuarine mud. \textit{Cont. Shelf Res.} \textbf{2002},
\textit{22(9)}, 1339--1360.
\url{https://doi.org/10.1016/S0278-4343(02)00010-9}

\bibitem{Zaneveld2003} Zaneveld, J. R. V.; Pegau, W. S. Robust underwater
visibility parameter. \textit{Opt. Express} \textbf{2003}, \textit{11(23)},
2997--3009.
\url{https://doi.org/10.1364/OE.11.002997}

\end{thebibliography}
\end{document}